\documentclass[aps,pra,twocolumn,superscriptaddress]{revtex4-1}
\usepackage{graphicx}
\usepackage{dcolumn}% Align Table columns on decimal point
\usepackage{bm}% bold math
\usepackage{multirow}
\usepackage{ulem}
\usepackage{color}
\usepackage{subfigure}
\usepackage{amsmath}
\usepackage{upgreek}
\usepackage{amsthm,amsmath,amssymb}
\usepackage{lineno}

\begin{document}

\preprint{APS/123-QED}

\title{Precise determination of ground-state hyperfine splitting and calculation of Zeeman coefficients for $^{171}$Yb$^+$ microwave frequency standard}

\author{J. Z. Han}
\affiliation{
State Key Laboratory of Precision Measurement Technology and Instruments, Key Laboratory of Photon Measurement and Control Technology of Ministry of Education, Department of Precision Instrument, Tsinghua University, Beijing 100084, China}%

\author{B. Q. Lu}
\affiliation{
National Time Service Center, Chinese Academy of Sciences, Xi’an 710600, China}%

\author{N. C. Xin}
\affiliation{
State Key Laboratory of Precision Measurement Technology and Instruments, Key Laboratory of Photon Measurement and Control Technology of Ministry of Education, Department of Precision Instrument, Tsinghua University, Beijing 100084, China}%

\author{Y. M. Yu}
\email{ymyu@aphy.iphy.ac.cn}
\affiliation{Beijing National Laboratory for Condensed Matter Physics, Institute of Physics, Chinese Academy of Sciences, Beijing 100190, China}
\affiliation{University of Chinese Academy of Sciences, Beijing 100049, China}

\author{H. R. Qin}
\affiliation{
Department of Physics, Tsinghua University, Beijing 100084, China}%

\author{S. T. Chen}
\affiliation{
State Key Laboratory of Precision Measurement Technology and Instruments, Key Laboratory of Photon Measurement and Control Technology of Ministry of Education, Department of Precision Instrument, Tsinghua University, Beijing 100084, China}%

\author{Y. Zheng}
\affiliation{
Department of Physics, Tsinghua University, Beijing 100084, China}%

\author{J. G. Li}
\email{li\_jiguang@iapcm.ac.cn}
\affiliation{
Institute of Applied Physics and Computational Mathematics, Beijing 100088, China}%

\author{J. W. Zhang}
\email{zhangjw@tsinghua.edu.cn}
\affiliation{
State Key Laboratory of Precision Measurement Technology and Instruments, Key Laboratory of Photon Measurement and Control Technology of Ministry of Education, Department of Precision Instrument, Tsinghua University, Beijing 100084, China}%

\author{L. J. Wang}
\email{lwan@mail.tsinghua.edu.cn}
\affiliation{
State Key Laboratory of Precision Measurement Technology and Instruments, Key Laboratory of Photon Measurement and Control Technology of Ministry of Education, Department of Precision Instrument, Tsinghua University, Beijing 100084, China}%
\affiliation{
Department of Physics, Tsinghua University, Beijing 100084, China}%

\date{\today}% It is always \today, today,
             %  but any date may be explicitly specified

\begin{abstract}
We report precise measurement of the hyperfine splitting and calculation of the Zeeman coefficients of the $^{171}$Yb$^+$ ground state. The absolute hyperfine splitting frequency is measured using high-resolution laser-microwave double-resonance spectroscopy at 0.1 mHz level, and evaluated using more accurate Zeeman coefficients. These Zeeman coefficients are derived using Land\'{e} $g_J$ factors calculated by two atomic-structure methods, multiconfiguration Dirac-Hartree-Fock, and multireference configuration interaction. The cross-check of the two calculations ensures an accuracy of the Zeeman coefficients at $10^{-2}$ Hz/G$^2$ level. The results provided in this paper improve the accuracy and reliability of the second-order Zeeman shift correction, thus further improving the accuracy of the microwave frequency standards based on $^{171}$Yb$^+$. The high-precision hyperfine splitting and Zeeman coefficients could also support could also support further experiments to improve the constraints of fundamental constants through clock frequency comparison of the Yb$^+$ system.
\end{abstract}

%\keywords{Suggested keywords}%Use showkeys class option if keyword
                              %display desired
\maketitle

%\tableofcontents

\section{Introduction}
The microwave frequency standard is one of the most widely used quantum technologies. Those based on trapped ions are considered the next generation of atomic frequency standards due to their potentially high stability, high accuracy, and transportability \cite{schmittberger2020review}. Such frequency standards are expected to promote development in areas such as satellite navigation \cite{bandi2011high, mallette2010space}, deep space exploration \cite{prestage2007atomic, liu2018orbit, burt2021demonstration}, timekeeping \cite{diddams2004standards, burt2008compensated} and telecommunications \cite{ho1998new}. At present, trapped-ion microwave frequency standards are mostly based on $^{199}$Hg$^+$ \cite{berkeland1998laser, burt2021demonstration}, $^{113}$Cd$^+$ \cite{tanaka1996determination, jelenkovic2006high, qin2022high} and $^{171}$Yb$^+$ \cite{phoonthong2014determination, mulholland2019laser, xin2022laser}. The clock transition of those candidates is their ground-state hyperfine splittings.

$^{171}$Yb$^+$ ion, which benefits from a simple structure ($I=1/2$) and relatively large ground-state hyperfine splitting ($\nu_{\rm HFS}=12.6$-GHz), is very suitable for the development of microwave frequency standards. More importantly, lasers for cooling and repumping $^{171}$Yb$^+$ ions can be easily obtained by compact semiconductor lasers and are fiber-friendly. The above advantages make the laser-cooled $^{171}$Yb$^+$ scheme stand out among other candidates for trapped-ion microwave frequency standards for practical use. In addition, the $^{171}$Yb$^+$ ion is also widely used in areas such as optical frequency standards \cite{schneider2005sub, huntemann2016single}, quantum computation \cite{lu2019global, morong2021observation, yang2022realizing}, and searching for new physics based on trapped ions \cite{godun2014frequency, huntemann2014improved, counts2020evidence, lange2021improved}. Several metrology laboratories have focused on developing microwave frequency standards based on trapped $^{171}$Yb$^+$ ions. For instance, the National Institute of Information and Communications in Japan has developed a laser-cooled $^{171}$Yb$^+$ ion microwave frequency standard with a stability of $2.1\times10^{-12}/\sqrt{\tau}$ and an evaluated accuracy of $1.1\times10^{-14}$ \cite{phoonthong2014determination}. The National Physical Laboratory in the UK has built a prototype with a stability of $3.6\times10^{-12}/\sqrt{\tau}$. The entire system fits into a 6U 19-inch rack unit ($51\times49\times28$) cm$^3$, revealing its potential for miniaturization \cite{mulholland2019compact, mulholland2019laser}. The Sandia National Laboratories, the Jet Propulsion Laboratory, and the National Institutes of Standards and Technology in the USA have developed a chip-scaled $^{171}$Yb$^+$ microwave frequency standard. The vacuum system is several cm$^3$ in size and achieved a stability of $2\times10^{-11}/\sqrt{\tau}$ \cite{schwindt2015miniature, schwindt2016highly}. Microchip Technology in the USA has developed a buffer gas-cooled $^{171}$Yb$^+$ microwave frequency standard for military and commercial applications \cite{park2020171}. The National Measurement Laboratory in Australia has also predicted a frequency stability of better than $5\times10^{-14}$ and a frequency uncertainty of $4\times10^{-15}$ for a laser-cooled $^{171}$Yb$^+$ microwave frequency standard \cite{park2007171Yb+}.

Our group at Tsinghua University has been committed to developing microwave frequency standards based on laser-cooled Cd$^+$ \cite{zhang2012high, wang2013high, miao2015high, zuo2019direct, han2019theoretical, han2019roles, han2021toward, miao2021precision, han2022isotope, han2022determination, qin2022high} and Yb$^+$ \cite{xin2022laser} for the past fifteen years. In particular, we have achieved the most stable laser-cooled $^{171}$Yb$^+$ microwave frequency standard ($8.5\times10^{-13}/\sqrt{\tau}$) and aim to evaluate its accuracy to the $10^{-15}$ level \cite{xin2022laser}. A high accuracy, transportable $^{171}$Yb$^+$ microwave frequency standard can provide a better time-frequency reference for the China's BeiDou Navigation Satellite System (BNSS). However, to achieve such a high accuracy, all frequency shifts of the clock transition need to be measured or evaluated down to 0.1 mHz level accuracy. These pose a challenge to both experimentalists and theorists.

Accurate frequency comparison of different clock transitions between atomic levels of the electronic or hyperfine transitions over time also can be used to search for new boson \cite{counts2020evidence}, to test local position invariance \cite{lange2021improved}, and to investigate variations of the fine structure constant $\alpha$ \cite{godun2014frequency} and the proton-to-electron mass ratio $\mu$ \cite{huntemann2014improved}. For instance, constraints of $\dot{\alpha}/\alpha<-0.20(20)\times10^{-16}$/yr and $\dot{\mu}/\mu<-0.5(1.6)\times10^{-16}$/yr are given using frequency comparisons of the $^{171}$Yb$^+$ 467-nm E3 optical transition and $^{133}$Cs 9.2-GHz ground-state hyperfine splitting \cite{huntemann2014improved}. The constraints may be further improved by using frequency comparison of the $^{171}$Yb$^+$ 467-nm optical clock transition and $^{171}$Yb$^+$ 12.6-GHz ground-state hyperfine splitting, as some common-mode noise can be suppressed. Such frequency comparison also relies on accurately evaluating the absolute clock frequencies, especially the 12.6-GHz ground-state hyperfine splitting frequency in $^{171}$Yb$^+$.

Among all the systematical energy shifts of the measurement of the the clock frequency, i.e., the ground-state hyperfine splitting, the second-order Zeeman shift caused by the magnetic field is dominant \cite{berkeland1998laser, phoonthong2014determination, qin2022high}. Note that the second-order Zeeman shift of in a trapped-ion microwave frequency standard is approximately four orders of magnitude larger than other shifts, even in a strong magnetic shielding environment composed of five layers of $\mu$-mental and one layer of soft iron \cite{miao2021precision, qin2022high}. Therefore, the second-order Zeeman shift must be carefully evaluated. For the $6s~^2S_{1/2}~(F=0)\rightarrow(F=1)$ 12.6-GHz hyperfine splitting, the second-order Zeeman shift $\Delta \nu_{\rm SOZS}$ can be expressed as \cite{itano2000external},
\begin{equation}
\Delta \nu_{\rm SOZS}=K_0 B_0^2=\frac{(g_J-g'_I)^2 \mu_B^2}{2 h^2 \nu_{\rm HFS}} B_0^2,
\label{eq:K0}
\end{equation}
where $K_0=(g_J-g'_I)^2 \mu_B^2/(2 h^2 \nu_{\rm HFS})$ is the second-order Zeeman coefficient, $\nu_{\rm HFS}$ is the ground-state hyperfine splitting, $g_J$ and $g'_I$ are the electronic and nuclear $g$ factor, $h$ is the Planck constant and $\mu_B$ is the Bohr magneton. Therefore, the accuracy of $K_0$ is mainly limited by the accuracy of $g_J$ since $\nu_{\rm HFS}$ and $g'_I$ are accurate enough. However, there are only two theoretical results and one early measurement of the ground-state Land\'{e} $g_J$ factor in $^{171}$Yb$^+$ and the results are not consistent: one giving 2.002798(113), calculated by relativistic-coupled-cluster (RCC) theory \cite{yu2020ground}; another giving 2.003117 without specifying uncertainty, calculated by time-dependent Hartree-Fock (TDHF) method \cite{gossel2013calculation}; and the other giving 1.998 without specifying uncertainty, an early spectroscopic measurement \cite{meggers1967second}. Even for two theoretical results, there exists a difference of around 0.0003 that generates a fractional second-order Zeeman shift of about $(0.8\sim 7.9)\times10^{-14}$ in a typical static magnetic field ($0.03\sim0.10$ G) \cite{berkeland1998laser, phoonthong2014determination, qin2022high} for operating trapped-ion microwave frequency standard. The large uncertainty of second-order Zeeman shift caused by the larger error bar of the Land\'{e} $g_J$ factor can obviously deteriorate the accuracy of the latest $^{171}$Yb$^+$ microwave frequency standard of our group (goal $<9\times10^{-15}$ \cite{xin2022laser}) and others ($1.1\times10^{-14}$ \cite{phoonthong2014determination}).

In this work, we report on the precise determination of the ground-state hyperfine splitting and calculation of Zeeman coefficients for $^{171}$Yb$^+$ microwave frequency standard. The hyperfine splitting is measured using laser-microwave double-resonance spectroscopy \cite{vogel2018laser} in our laser-cooled $^{171}$Yb$^+$ microwave frequency standard and evaluated using the new Zeeman coefficients obtained in this paper. Those coefficients are derived using the Land\'{e} $g_J$ factor calculated by two atomic structure calculation methods in this paper: the multiconfiguration Dirac-Hartree-Fock (MCDHF) and the multireference configuration interaction (MRCI). The cross-check of the new Zeeman coefficients guarantee clock transition evaluation of $10^{-15}$ accuracy under typical conditions. The results reported in this paper are of great importance for further improving the performance of the microwave frequency standards based on trapped $^{171}$Yb$^+$ ions. Our efforts on the ground state Land\'{e} $g_J$ factor and hyperfine splitting could also support further experiments in improving the constraints of fundamental constants through clock frequency comparison of the Yb$^+$ system.

\section{Measurement of hyperfine splitting $\nu_{\rm HFS}$}

\begin{figure*}
\centering
\resizebox{0.95\textwidth}{!}{
\includegraphics{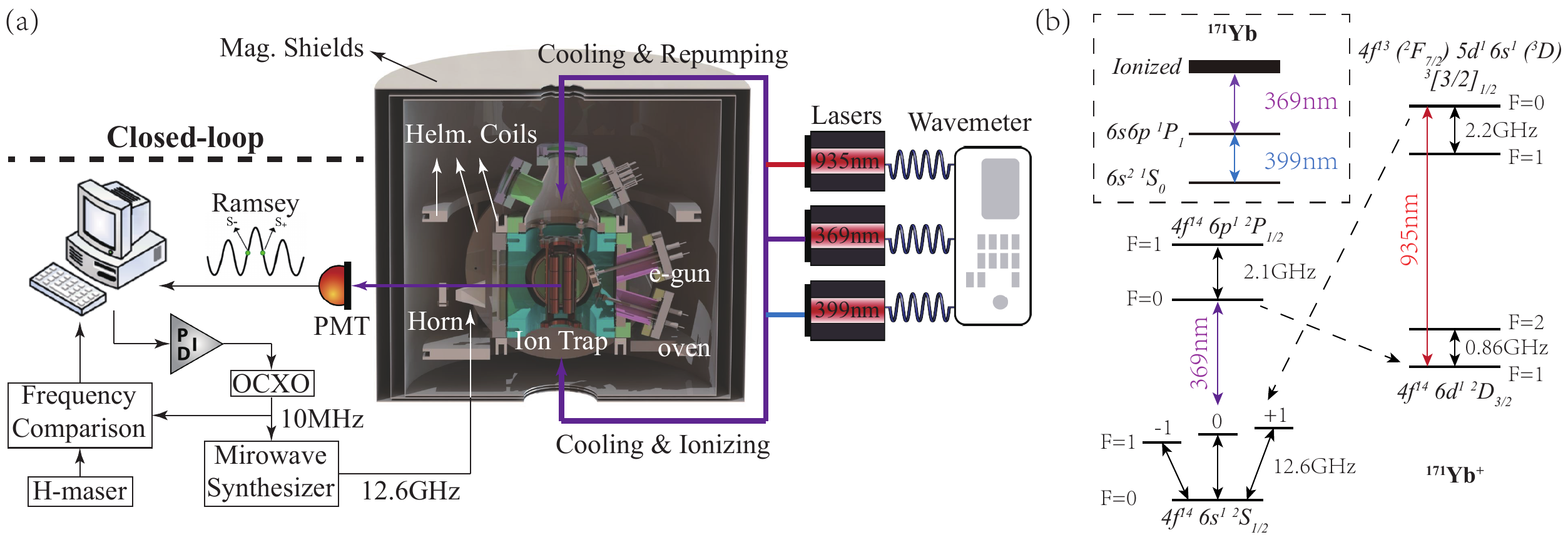}
}
\caption{(a) Schematic of the experimental setup for measuring the hyperfine splitting $\nu_{\rm HFS}$ of $^{171}$Yb$^+$. Lasers with wavelengths of 369 nm, 399 nm, and 935 nm are used for cooling \& probing, ionizing, and repumping of $^{171}$Yb$^+$. $\nu_{\rm HFS}$ is measured under a closed-loop operation. PMT, photomultiplier tubes; OCXO, oven-controlled crystal oscillator; PID, proportional-integral-differentiation. (b) Schematic energy levels of the $^{171}$Yb$^+$ ion and $^{171}$Yb atom (not to scale).}
\label{fig:exp}
\end{figure*}

\begin{figure}
\centering
\resizebox{0.45\textwidth}{!}{
\includegraphics{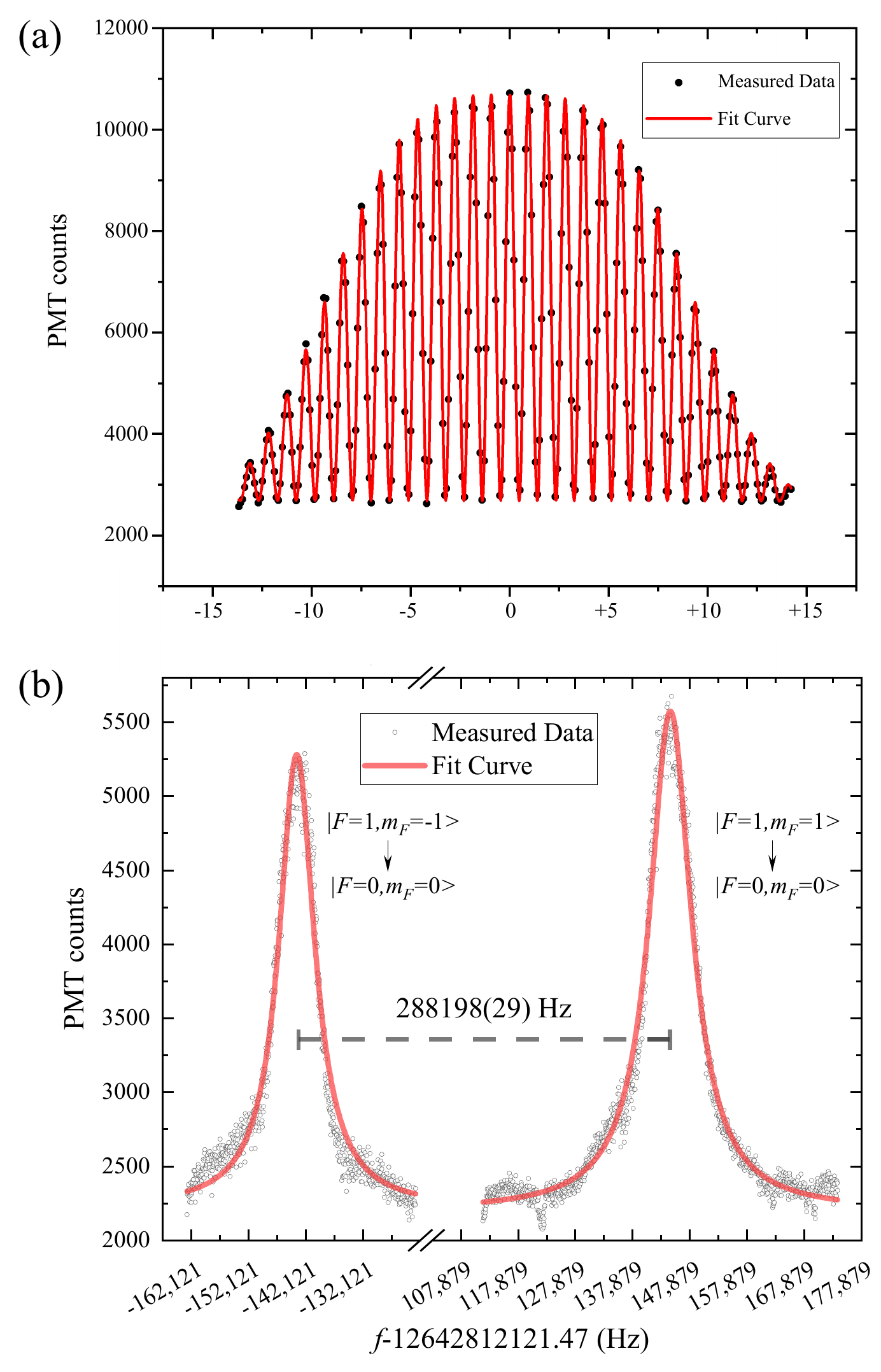}
}
\caption{(a) Typical Ramsey fringe of the clock transition (12.6-GHz) with a free evolution time of 1000 ms, a $\pi/2$ microwave pulse time of 60 ms, a microwave power of -29.8 dBm, and a fluorescence signal integration time of 150 ms. The central frequency and linewidth of the Ramsey fringe are 12642812121.47 Hz and 434 mHz. (b) Rabi fringe of the Zeeman transition with a $\pi$ microwave pulse time of 60 ms, microwave power of -3 dBm, and scan step of 400 Hz. Peaks from left to right are $F=1,m_F=-1\rightarrow F=0,m_F=0$ and $F=1,m_F=1\rightarrow F=0,m_F=0$ transitions, respectively.}
\label{fig:result}
\end{figure}

The hyperfine splitting $\nu_{\rm HFS}$ is measured through a laser-microwave double-resonance spectroscopy in our laser-cooled $^{171}$Yb$^+$ microwave frequency standard. The experiment is conducted in a linear Paul radio-frequency (RF) ion trap, and the schematic of the setup is shown in Fig. \ref{fig:exp} (a). The ratio of $R=r_0$ is optimized and set at 1.1468, for which $2R=14.22$ mm is the outer diameter of the electrode and $r0=6.2$ mm is the radial distance from the axis of the trap to the closest surface of the electrodes, to reduce heating effects from higher-order RF potential and to increase the number of trapped ions. A magnetic shield of three layers of $\mu$-metal is installed outside the vacuum chamber to shield the external magnetic field. Approximately $10^5$ $^{171}$Yb$^+$ ions are loaded in the ion trap for hyperfine splitting measurement, and the energy levels of $^{171}$Yb$^+$ are shown in Fig. \ref{fig:exp} (b). The ions are photon-ionized from $^{171}$Yb atoms by a 399-nm (Toptica DLpro, $6s^2~^1S_0\rightarrow 6s6p~^1P_1$) and a 369-nm (Precilasers FL-SF, $6s6p~^1P_1\rightarrow$ ionized continuum) laser beam. The temperature of ions is reduced from laser Doppler cooling by a 369-nm ($6s~^2S_{1/2}\rightarrow 6p~^2P_{1/2}$) laser beam to reduce the second-order Doppler shift. A 935-nm laser beam (Toptica DLpro, $6d~^2D_{3/2}\rightarrow 5d6s~^3[3/2]_{1/2}$) and 12.6-GHz microwave radiation (Agilent E8257D, $6s~^2S_{1/2}~F=0\rightarrow 6s~^2S_{1/2}~F=1$) are applied to repump the ions from $^2D_{3/2}$ and $^2S_{1/2}~(F=0)$ dark states back to the cooling cycle. The frequency of each laser beam is measured and stabilized using a high-precision wavemeter (HighFinesse WS8-2). Three pairs of Helmholtz coils installed near the vacuum chamber are used to provide the quantization axis (several $10^{-2}$ G) of the $^{171}$Yb$^+$ ions and compensate the geomagnetic field. Another pair of Helmholtz coils generate a 7-G strong magnetic field. The direction of the strong magnetic field magnetic field forms an angle of approximately $30^{\circ}$ with the 369-nm laser beam to induce a precession in the dipole moments of the ions and to destabilize the Zeeman dark states of the $^2S_{1/2}~(F=1)$ level \cite{berkeland2002destabilization}.

After the ions are loaded and cooled, the hyperfine splitting $\nu_{\rm HFS}$ is measured by Ramsey’s method of separated oscillatory fields in a closed-loop operation. Large ion cloud in a linear ion trap suffers from RF heating, unlike single ion or ion string. The optimized ion trap ensures that up to $10^5$ $^{171}$Yb$^+$ can be stably trapped in the ion trap at a relatively low temperature during the clock signal's interrogation. In the passive quantum frequency standard, the signal-to-noise ratio (SNR) is proportional to $\sqrt{N}$, where $N$ is the number of quantum particles. Thus, $10^5$ of ions ensures their clock transition can be detected with a high SNR. A typical, high SNR Ramsey fringe of the clock transition detected by photomultiplier tubes (PMT, Hamamatsu H12386-210) is shown in FIG. \ref{fig:result} (a), where the full width at half maximum (FWHM) of the center fringe is $2\delta=434$ mHz. The output frequencies of the oven-controlled crystal oscillator (OCXO, Rakon HSO14) are locked by the central peak of the Ramsey fringe through a proportional-integral-differentiation (PID) controller and are recorded by comparison with the H-maser. The clock transition under a static magnetic field $\nu_{\rm HFS}(B_0)$ can be calculated by using
\begin{eqnarray}
\nu_{\rm clock}(B_0)&=&f_{\rm OCXO}\times M_{\rm Synthesizer}
\label{eq:A}
\end{eqnarray}
where $f_{\rm OCXO}$ is the average value of the output frequency of the OCXO, $M_{\rm Synthesizer}$ is the magnification of the microwave synthesizer.

The static magnetic field $B_0$ experienced by the ions can be calibrated by measuring the Larmor frequency difference $\Delta \nu_L$ of the $6s~^2S_{1/2}~(F=0,m_F=0)\rightarrow(F=1,m_F=\pm 1)$ Zeeman sublevels. The Larmor frequency is measured by using the Rabi oscillation spectra, as shown in Fig. \ref{fig:result} (b). The static magnetic field $B$ can be calculated by using
\begin{equation}
B_0=\frac{\Delta \nu_L}{(g_J+g'_I) \mu_B/h}=\frac{\Delta \nu_L}{2 K_Z},
\label{eq:Kz}
\end{equation}
where $K_Z=(g_J+g'_I) \mu_B/(2h)$ is the first-order Zeeman coefficient. The second-order Zeeman shift of the hyperfine splitting can be evaluated by using Eq. (\ref{eq:K0}). The absolute hyperfine splitting $\nu_{\rm HFS}$ can be derived by $\nu_{\rm HFS}(B_0)$ subtracting the second-order Zeeman shift and other systematic frequency shifts.

\section{THEORY}

\subsection{Operator of Land\'{e} $g_J$ factor}
The interaction Hamiltonian between an atom and the magnetic field can be written as \cite{andersson2008hfszeeman}
\begin{eqnarray}
H_m=(\boldsymbol{N^{(1)}}+\boldsymbol{\Delta N^{(1)}})B,
\end{eqnarray}
by choosing the direction of the magnetic field as the $z$-direction and neglecting all diamagnetic contributions in a relativistic frame. The electronic tensor operators of an N-electron atom can be expressed as \cite{cheng1985ab},
\begin{eqnarray}
    \boldsymbol{N^{(1)}}&=&\sum^N_{i=1}\boldsymbol{n^{(1)}}(i)=\sum^N_{i=1}-I\frac{\sqrt{2}}{2\alpha}r_i(\boldsymbol{\alpha_i}\boldsymbol{C^{(1)}}(i))^{(1)},\nonumber\\
    \boldsymbol{\Delta N^{(1)}}&=&\sum^N_{i=1}\boldsymbol{\Delta n^{(1)}}(i)=\sum^N_{i=1}\frac{g_s-2}{2}\beta_i\boldsymbol{\Sigma_i},
\end{eqnarray}
where $\Sigma_i$ is the relativistic spin-matrix, $I$ is the imaginary unit, and $g_s=2.00232$ is the g factor of the electron spin corrected for QED effects. The $\boldsymbol{\Delta N^{(1)}}$ term is the Schwinger QED correction. The interaction Hamiltonian $H_m$ can be treated in first-order perturbation theory in a weak magnetic field situation. A fine-structure level $\Gamma J$ is split according to
\begin{eqnarray}
    &&\langle \Gamma J M_J | N_0^{(1)}+\Delta N_0^{(1)} | \Gamma J M_J \rangle B \nonumber\\
    &=&\frac{M_J}{\sqrt{J(J+1)}}\langle \Gamma J || N_0^{(1)}+\Delta N_0^{(1)} || \Gamma J \rangle B \nonumber\\
    &=&g_J M_J \frac{B}{2},
\end{eqnarray}
where $J$ is total angular momentum, $M_J$ is component along the $z$ direction of $J$, $\Gamma$ represent other appropriate angular momentum, and Land\'{e} $g_J$ is defined as
\begin{equation}
    g_J=2\frac{\langle \Gamma J || N_0^{(1)}+\Delta N_0^{(1)} || \Gamma J \rangle}{\sqrt{J(J+1)}}.
\end{equation}

\subsection{Multiconfiguration Dirac-Hartree-Fock approach}

The MCDHF method \cite{jonsson2022introduction}, as implemented in the {\sc Grasp} package \cite{jonsson2023grasp}, is employed to obtain wave functions referred to as the atomic state function (ASF). The ASF are approximate eigenfunctions of the Dirac Hamiltonian describing a Coulombic system,
\begin{eqnarray}
H_{\rm DC}&=&\sum_i [c(\vec { \bm{\alpha}}\cdot \vec {\bf p})_i+(\bm \beta_i-1) c^2+V_{nuc}(r_i) ] +\sum_{i<j}^N\frac{1}{r_{ij}}\nonumber\\
&&-\frac{1}{2r_{ij}}[\vec { \bm{\alpha_i}}\cdot \vec { \bm{\alpha_j}}+\frac{(\vec { \bm{\alpha_i}}\cdot \vec { \bm{r_{ij}}})(\vec { \bm{\alpha_j}}\cdot \vec { \bm{r_{ij}}})}{r_{ij}^2}],
\end{eqnarray}
where $ \vec {\bm \alpha}$ and $\bm \beta$ represent the Dirac matrices, $\vec {\bf p}$ is the momentum operator, $r_{ij}$ is the distance between electrons $i$ and $j$, and $V_{nuc}(r)$ is the nuclear potential results from a nuclear charge density given by a two-parameter Fermi distribution function \cite{parpia1992relativistic}. The last term is the Breit interaction in the low-frequency approximation. The ASF is a linear combination of configuration state functions \cite{grant2007relativistic},
\begin{eqnarray}
\Psi(\Gamma J M_J)=\sum^{N_{\rm CSF}}_{i=1}c_i\Phi(\Gamma_i J M_J),
\end{eqnarray}
where $c_i$ represents the mixing coefficient corresponding to the $i$th configuration state function. The CSFs are the linear combinations of one-electron Dirac orbital products. 

\begin{figure}
\centering
\resizebox{0.49\textwidth}{!}{
\includegraphics{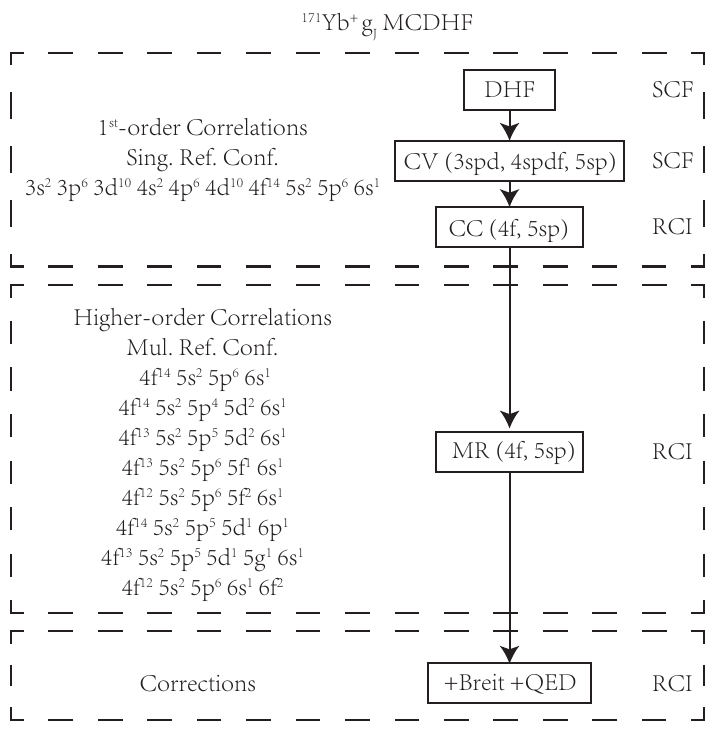}
}
\caption{Computational model of the MCDHF method. The model takes into account the main first-order and higher-order electron correlations. The first-order electron correlations include the CV and CC correlations. Corrections for the Breit and QED effects are added in the final result.}
\label{fig:mcdhf}
\end{figure}

The active space approach is adopted to capture electron correlations in this work. Our computational model is shown in FIG. \ref{fig:mcdhf}. The computation started from the Dirac-Hartree-Fock (DHF) approximation. The occupied spectroscopic orbitals are optimized in the single reference configuration $\{[\rm Ne]3s^23p^63d^{10}4s^24p^64d^{10}4f^{14}5s^25p^6~6s\}$. According to perturbation theory, electron correlations can be divided into first-order and higher-order correlations. The outermost $6s$ orbitals in the reference configuration are treated as the valence electrons, and the others are treated as the core. The correlations between the valence and core electrons (CV) are considered in the self-consistent field (SCF) calculations. The CV correlations are taken into account using the single and restricted double (SrD) substitutions from the $n\ge 3$ core orbitals. The SrD substitutions mean that only one electron in each core orbital can be excited to the active set. The active set is expanded layer by layer to $\{14s,13p,12d,12f,10g,10h,8i\}$, and only the added correlation orbitals are optimized each time. The correlations between core electrons (CC) are considered in the RCI calculations where only the mixing coefficients are variable. The CC correlations are accounted by CSFs generated through SD subsitituions from allowing the single and double (SD) substitutions from the $4f,5s,5p$ core orbitals to the largest active set. The main higher-order correlations are captured by the multireference (MR) SD-excitation method \cite{li2012effects, filippin2016multiconfiguration, bieron2009complete}. The MR configuration set was formed by selecting the dominant CSFs in the CC model, i.e., those CSFs with mixing coefficients $c_i$ larger than 0.02. Finally, the Breit interaction and QED (vacuum polarization and self-energy) are considered in the RCI procedure. CSFs that do not interact with the reference configurations are removed to raise the computational efficiency \cite{jonsson2007grasp2k, froese2022computational}.

\subsection{Multireference Configuration Interaction approach}

The calculation of the Yb$^{+}$ ion is carried out by using the multireference configuration interaction (MRCI) method that is based on the general active space (GAS) \cite{fleig2001generalized, fleig2003generalized, fleig2006generalized} and implemented in the KR-CI module of the Dirac package \cite{knecht2008large, jensen_h_j_aa_2022_6010450}. We start with the close-shell Dirac-Hartree-Fock (DHF) calculation of Yb$^{2+}$ under the Dirac-Coulomb-Gaunt Hamiltonian. The Dirac-Coulomb-Gaunt Hamiltonian is written as
\begin{eqnarray}
H_{\rm DCG}&=&\sum_i [c(\vec { \bm{\alpha}}\cdot \vec {\bf p})_i+(\bm \beta_i-1) c^2+V_{nuc}(r_i) ] + \nonumber \\
&& \sum_{i<j}\bigg[\frac{1}{r_{ij}}-\frac{ {\vec {\bm \alpha}}_i \cdot {\vec { \bm \alpha}}_j }{2r_{ij}}\bigg]. \label{DCG}
\end{eqnarray}
The last term in Eq. (\ref{DCG}) represents the Gaunt interaction, which is the leading term of the Breit interaction.
	
In the DHF calculation, a Gaussian basis set of quadruple-zeta quality, dyall.cv4z, is adopted \cite{gomes2010relativistic}. Next, the electron correlation calculation is performed based on the optimized spinors that are obtained by the DHF calculation. We correlate the outermost 23 electrons of Yb$^{+}$, which takes into account the single (S) excitation from the $5s$ and $5p$ shells, the single and double (SD) excitation from the $4f$ shell and the single, double and triple (SDT) excitation from the $6s$ and $5d$ valent shells. The virtual energy orbitals with energy above 10 a.u. are truncated off from the configuration interaction calculation, considering their negligible impact on the MRCI results. The DHF and MRCI calculations are carried out under the Dirac package. 

\section{Results \& Discussion}

\begin{table}
\caption{Land\'{e} $g_J$ factor, first-order Zeeman coefficient ($K_Z$, in $10^3$ Hz/G), second-order Zeeman coefficient ($K_0$, in Hz/G$^2$) and hyperfine splitting ($\nu_{\rm HFS}$, in mHz) of ground-state $^{171}$Yb$^+$ obtained in this work. The $K_0$ coefficients are calculated based on $\nu_{\rm HFS}=12642812118.47(1)$ Hz. Results of other works are also listed for comparison.} \label{tab:1}
{\setlength{\tabcolsep}{4pt}
\begin{tabular}{llllllll}\hline\hline
\multicolumn{4}{c}{Land\'{e} $g_J$ factor} \\
 \multicolumn{2}{l}{$g_J$} & Source & \\\hline
 \multicolumn{2}{l}{2.002630(38)}  & MCDHF (This paper) & Theor.\\
 \multicolumn{2}{l}{2.002604(55)} & MRCI (This paper) & Theor.\\
 \multicolumn{2}{l}{2.002617(68)}  & Final (This paper) & Theor.\\
 \multicolumn{2}{l}{2.002798(113)}  & RCC \cite{yu2020ground}  & Theor.\\
 \multicolumn{2}{l}{2.003117}  & TDHF \cite{gossel2013calculation} & Theor. \\
 \multicolumn{2}{l}{2.0023}  & Nonrelat. \cite{vanier1989quantum} & Theor. \\
 \multicolumn{2}{l}{1.998}  & Spectr. \cite{meggers1967second} & Expt. \\
 \\
 \multicolumn{4}{c}{Zeeman coefficients} \\
 $K_0$ & $K_Z$ & Source & \\\hline
 310.874(12) & 1401.089(27) &  MCDHF (This paper) & Theor.\\
 310.866(17) & 1401.071(38) & MRCI (This paper) & Theor.\\
 310.870(21) & 1401.080(47) &  Final (This paper) & Theor.\\
 310.77 & 1400.86 & Nonrelat. \cite{vanier1989quantum} & Theor.\\
 310.93(35) & 1401.206(79) & RCC \cite{yu2020ground}$^a$  & Theor. \\
 311.03 & 1401.430 & TDHF \cite{gossel2013calculation}$^a$ &Theor.\\
 309$^a$ & 1398$^a$ & Spectr. \cite{meggers1967second}$^a$ & Expt. \\
 \\
 \multicolumn{4}{c}{Hyperfine splitting} \\
 \multicolumn{2}{l}{$\nu_{\rm HFS}$} & Source & \\\hline
 \multicolumn{2}{l}{12642812118468.9(8)}  & This paper & Expt. \\
 \multicolumn{2}{l}{12642812118468.2(4)}  & \cite{phoonthong2014determination} & Expt. \\
 \multicolumn{2}{l}{12642812118468.5(9)} &  \cite{warrington2002microwave} & Expt. \\
 \multicolumn{2}{l}{12642812118466(2)} &  \cite{fisk1995performance} & Expt. \\
 \multicolumn{2}{l}{12642812118471(9)} &  \cite{tamm1995radio} & Expt. \\
 \multicolumn{2}{l}{12642812118468(16)} &  \cite{sellars1995further} & Expt. \\
\hline\hline
\multicolumn{4}{l}{$^a$Derived by using the Land\'{e} $g_J$ factors given in the Refs.} \\
\end{tabular}}
\end{table}

\subsection{Land\'{e} $g_J$ factor \& Zeeman coefficients}

The Land\'{e} $g_J$ factor, the first- and second-order Zeeman coefficient, and the hyperfine splitting of the $^{171}$Yb$^+$ ground-state obtained in this work are shown in TABLE \ref{tab:1}. The Land\'{e} $g_J$ factors calculated by the MCDHF and MRCI methods are in excellent agreement, and the deviation between the two is at the fifth decimal place. Uncertainties of Land\'{e} $g_J$ factors are given in parentheses. Accurate calculations of Land\'{e} $g_J$ factors have proven to be complicated even for alkali-metal atoms and alkali-metal-like ions \cite{sahoo2017relativistic}, as they are sensitive to electron correlations effects. Such effects of the electrons from the $4f$ orbitals and higher-order corrections in Yb$^+$ are also proved to be large and complicated. The inconsistency between the theory and experiment of the quadruple moment in Yb$^+$ $4f~^2F_{7/2}$ state is also due to the above reason \cite{nandy2014quadrupole}. These two results are within the uncertainty of the previous RCC result \cite{yu2020ground} and improve the accuracy of the Land\'{e} $g_J$ factor from the fourth to the fifth decimal place. All three results have a deviation of approximately 0.0004 from the previous results calculated by the TDHF method \cite{gossel2013calculation}. The theoretical results have a deviation up to 0.005 compared with the early spectroscopic result given by Meggers \cite{meggers1967second}. The anomaly correction of the Land\'{e} $g_J$ factor has the opposite sign compared with the theoretical results. A recent measurement of the second-order Zeeman coefficient of the $^2D_{5/2}$ in $^{171}$Yb$^+$ also notwithstanding with the early spectroscopic result \cite{tan2021precision}. Those suggests potential issues in the early
spectroscopic result of the Land\'{e} $g_J$ factor in Yb$^+$. Comparing the results of the same physical constant from different calculation methods is also crucial for investigating the role of electronic correlations and for developing atomic structure calculation theory. Therefore, we encourage more experimental and theoretical research on the Land\'{e} $g_J$ factors in Yb$^+$.

The $K_0$ and $K_Z$ coefficients of $^{171}$Yb$^+$ ground-state are calculated using Eq. (\ref{eq:K0}) and Eq. (\ref{eq:Kz}), where $g_J=2.002617(68)$ is the Land\'{e} $g_J$ factor recommended in this paper, $\nu_{\rm HFS}=12642812118.47(1)$ Hz is the hyperfine splitting, and $g'_I=-\mu_{\rm Yb} \cdot \mu_N/(I \cdot \mu_B)=-0.53772(1)\times10^{-3}$ is the nuclear g-factor of $^{171}$Yb$^+$ \cite{olschewski1972messung, tiesinga2021codata}. The uncertainties of those coefficients are given in parentheses. The widely used values of $K_0$ and $K_Z$ are $310.77$ Hz/${\rm G}^2$ and $1400.86\times10^3$ Hz/${\rm G}$ standards without specifying uncertainty for quantum frequency standard community \cite{vanier1989quantum}. Those coefficients are calculated using a non-relativistic Land\'{e} $g_J$ factor of 2.0023. As seen from the previous analysis, those coefficients cannot meet the requirements of the state-of-the-art $^{171}$Yb$^+$ microwave frequency standards. The accuracy of the clock transition may be underestimated without considering the uncertainty of the Zeeman coefficients. The more accurate and reliable $K_0$ and $K_Z$ coefficients calculated in this work guarantee the evaluation accuracy of the $^{171}$Yb$^+$ microwave frequency standard. Zeeman coefficients calculated using Land\'{e} $g_J$ factor of other works are also listed in TABLE \ref{tab:1} for comparison.

\subsection{hyperfine splitting $\nu_{\rm HFS}$}

The ground-state hyperfine splitting $\nu_{\rm HFS}$ is determined in our high stability ($8.5\times10^{-13}/\sqrt{\tau}$) laser-cooled $^{171}$Yb$^+$ microwave frequency standard. The clock transition under a static magnetic field $\nu_{\rm HFS}(B_0)$ is calculated to be 12642812121757.4(4) mHz by using Eq. (\ref{eq:A}) where $f_{\rm OCXO}$ is 9999999083.85136 mHz and $M_{\rm Synthesizer}$ is 1264.2813280027. The value shown in parenthesis is the statistical uncertainty estimated from Allan deviation data at 1000 s \cite{xin2022laser}. The Larmor frequency difference of the Zeeman transition $\Delta \nu_L$ is measured to be 288198(29) Hz. The static magnetic field $B_0$ is then calibrated to be 0.102849(10) G using Eq. (\ref{eq:Kz}) with the measured $\Delta \nu_L$ and the calculated $K_Z$. Therefore, the second-order Zeeman shift $\Delta \nu_{\rm SOZS}$ is evaluated to be 3288.32(67) mHz using Eq. (\ref{eq:K0}) with the calculated $K_0$ in this paper. In addition, the reference shift caused by the frequency difference between the H-maser and the primary frequency standards is measured to be 0.555(6) mHz by the method of GPS common-view. The black-body radiation Zeeman shift (BBRZS) is estimated to be $-1.2\times10^{-4}$ mHz using the method in Ref. \cite{han2019theoretical}. Frequency shifts, including the second-order Doppler shift (SODS), the black-body radiation Stark shift (BBRSS), the quadratic Stark shift (QSS), and the gravitational Redshift (GRS) are estimated to be -0.517(40) mHz (see Ref. \cite{xin2022laser} for detail). Estimated systematic shifts and uncertainties are shown in TABLE \ref{tab:2}.

\begin{table}
\caption{Magnitudes (in mHz), uncertainties (Uncert., in mHz) and Fractional uncertainties (Frac., in $10^{-14}$) of the estimated systematic frequency shifts, hyperfine splitting under static magnetic field ($\nu_{\rm HFS}(B_0)$) and the estimated absolute hyperfine splitting ($\nu_{\rm HFS}$) in $^{171}$Yb$^+$. The uncertainty of $\nu_{\rm HFS}$ is the square root of the sum of the square of the uncertainty in $\nu_{\rm HFS}(B_0)$ and the uncertainty in total systematic shift. The uncertainties of the SOZS stem from $B_0$ and $K_0$ are listed separately for comparison. The predicted uncertainties of the SOZS in the improved system are also listed.} \label{tab:2}
{\setlength{\tabcolsep}{2pt}
\begin{tabular}{lllll}\hline\hline
Item & Magnitude & Uncert. & Frac. & Source \\\hline
SOZS (Current) &  3288.32 & 0.64 & 5.06 &$B_0$\\
 ($0.1\pm10^{-5}$ G) &   & 0.22 & 1.74 &$K_0$\\
SOZS (Improved) &   & 0.02 & 0.15 &$B_0$\\
 ($0.03\pm10^{-6}$ G) &  & 0.02 & 0.15  &$K_0$\\
Reference & 0.56 & 0.01 & 0.05\\
SODS & -0.44 & 0.04 & 0.30\\
GRS & 0.06 & 0.00 & 0.00\\
QSS & -0.01 & 0.00 & 0.00\\
BBRSS & -0.13 & 0.01 & 0.10\\
BBRZS & 0.0 & 0.00 & 0.00 \\
Pressure & 0.0 & 0.00 & 0.00 \\
Light & 0.0 & 0.00 & 0.00\\
Total & 3288.36 & 0.70 & 5.50\\
\\
$\nu_{\rm HFS}(B_0)$ & 12642812121757.4 & 0.4 & 3.0\\
$\nu_{\rm HFS}$ & 12642812118468.9 & 0.8  & 6.3 \\
\hline\hline
\end{tabular}}
\end{table}

Thus, the absolute frequency of the ground-state hyperfine splitting $\nu_{\rm HFS}$ of $^{171}$Yb$^+$ is determined to be 12642812118468.9(8) mHz by subtracting systematic frequency shifts, as listed in TABLE \ref{tab:2}. The uncertainty of $\nu_{\rm HFS}$ is the square root of the sum of the square of the uncertainty in $\nu_{\rm HFS}(B_0)$ and the uncertainty in total systematic shift. The measurement accuracy of $\nu_{\rm HFS}$ is mainly limited by the second-order Zeeman shift. The fractional uncertainties of the $\Delta \nu_{\rm SOZS}$ caused by the $B_0$ and $K_0$ are $5.06\times10^{-14}$ and $1.74\times10^{-14}$ respectively. The large fractional uncertainties mainly stem from the large value of the static magnetic field ($B_0\approx0.1$ G) in our experiment and the fluctuation of the external magnetic field ($\approx 10^{-5}$ G). The large $B_0$ value is due to the magnetization of the vacuum chamber by the strong magnetic field (7 G). Such magnetization can be mitigated using a nonmagnetic (e.g. Ti) vacuum chamber. The strong magnetic field can also be eliminated using a polarization-modulating scheme to remove hyperfine dark states. The $B_0$ value can be reduced to around $0.03$-G after those improvements \cite{berkeland1998laser, phoonthong2014determination}. In addition, the fluctuation of the external magnetic field can be further reduced to about $10^{-6}$ G after using more layers of $\mu$-metal \cite{han2021toward}. The fractional uncertainties of the $\Delta \nu_{\rm SOZS}$ caused by $B_0$ and $K_0$ can be reduced to about $1.5\times10^{-15}$ in such a situation. Those improvements are in progress.

The Zeeman coefficients calculated in this paper can guarantee the evaluation accuracy at the current and the improved situations. The high-precision $\nu_{\rm HFS}$ constant reported in this work is consistent with previous results. Our results not only support building a $10^{-15}$-level high-accuracy $^{171}$Yb$^+$ microwave frequency standard in China but also improve the reliability of $\nu_{\rm HFS}$ in $^{171}$Yb$^+$ ground-state and provide a good benchmark for the atomic structure calculations.

\section{CONCLUSION}

We report on the precise determination of ground-state hyperfine splitting and calculation of Zeeman coefficients for $^{171}$Yb$^+$ microwave frequency standard. The hyperfine splitting is measured using laser-microwave double-resonance spectroscopy in our laser-cooled $^{171}$Yb$^+$ microwave frequency standard. The first- and second-order Zeeman coefficients are derived using the recommended Land\'{e} $g_J$ factor in this paper. The Land\'{e} $g_J$ factor is calculated using the MCDHF and MRCI methods. The cross-check of the two methods ensures the reliability of the calculation results. The second-order Zeeman shift of $\nu_{\rm HFS}$ is evaluated using the measured Larmor frequency and the new Zeeman coefficients of this paper. The $\nu_{\rm HFS}$ constant is determined at the 0.1 mHz level and is consistent with previous results. The calculation and experiment conducted in this paper are significant for developing microwave frequency standards but also an excellent addition and benchmark to the current atomic structure calculations. The calculations and measurements are critical to building a high-performance, laser-cooled microwave frequency standard based on trapped $^{171}$Yb$^+$ ions in China. These results can also support further experiments to improve the constraints of fundamental constants through clock frequency comparison of the Yb$^+$ system.

\begin{acknowledgments}
The authors acknowledge R. Si, J. P. Liu, Z. M. Tang, and H. Wang for their helpful discussions. This work is supported by the National Key R\&D Program of China (No. 2021YFA1402100), the National Natural Science Foundation of China (No. 12073015), and the Space Application System of China Manned Space Program.

J.Z.H. and B.Q.L. contributed equally to this work.

\end{acknowledgments}

\bibliography{apssamp}% Produces the bibliography via BibTeX.

\end{document}